\documentclass[fleqn,usenatbib]{mnras}

\usepackage{newtxtext,newtxmath}
\usepackage[T1]{fontenc}
\usepackage{ae,aecompl} 

\usepackage{graphicx}   
\usepackage{amsmath}    

\newcommand{\Ha}{H${\alpha}$}
\newcommand{\Hb}{H${\beta}$}

\newcommand{\de}{$^\circ$}

\newcommand{\Msunpc}{$M_{\odot}\,pc^{-2}$}

\title[HI mass from optical spectroscopy]{The 
Neutral Hydrogen Mass in Galaxies Estimated via Optical Spectroscopy}

\author[Bizyaev]{
Dmitry Bizyaev$^{1,2}$\thanks{E-mail: dmbiz@apo.nmsu.edu}\\
$^{1}$Apache Point Observatory and New Mexico State University, Sunspot, NM, 88349, USA\\
$^{2}$Sternberg Astronomical Institute, Moscow State University, Universitetskiy prosp. 13, Moscow, 119234, Russia
}

\begin{document}

\date{Accepted 2023 December 1. Received 2023 November 19; in original form 2023 September 14.}

\pubyear{2023}

\label{firstpage}
\pagerange{\pageref{firstpage}--\pageref{lastpage}}
\maketitle

\begin{abstract}
We propose to employ emission line luminosities obtained via optical spectroscopy to estimate the 
content of neutral hydrogen (HI) in galaxies. We use the optical spectroscopy data from the 
Mapping of Nearby Galaxies at APO (MaNGA) survey released in the frames of public DR17 of the
Sloan Digital Sky Survey (SDSS). We compare the HI mass estimated by us for a large 
sample of SDSS/MaNGA galaxies with direct HI measurements from the 
ALFALFA survey and find a tight correlation between the masses with the correlation coefficient (CC) of 
0.91 and the r.m.s scatter of 0.15 dex for the logarithmic mass. 
The obtained relationship is verified via another sample of MaNGA galaxies with HI masses 
measured with the Green Bank Telescope. Despite the coarser angular resolution of the radio data, the relation
between the estimated and measured directly HI mass is tight as well - in this case CC=0.74
and the r.nm.s. is 0.29 dex. The established relations allow us to estimate the total mass 
of neutral hydrogen as well as the spatial distribution of HI surface
density in galaxies from optical spectroscopic observations only in a simple and efficient way.
\end{abstract}

\begin{keywords}
(galaxies:) intergalactic medium, galaxies: fundamental parameters, ISM: atoms, ISM: structure
\end{keywords}

\section{Introduction}
\noindent
Gas mass is a fundamental property of galaxies, which is connected to the star formation activity and history, galaxy assembly history, dust amount, and many other important characteristics. Studying the main gas phases – molecular, neutral and ionized, requires multi-wavelength observations. Traditional methods of measuring atomic gas mass employ radio observations, which have their advantages and disadvantages. For example, single-dish radio observations of HI have poor spatial resolution. 
Alternative methods of estimating the neutral hydrogen (HI) mass have been considered using e.g.
dust absorption \citep{brinchmann13} and optical spectra line ratio correlations \citep{stark21}
have been proposed. Correlations between optical colours and gas-to-stellar mass ratio is also 
widely used to probe the HI content in galaxies without performing radio 
observations \citep{kannappan04,zhang09,li10,eckert15}.

As shown by \citet{reynolds98}, the ratio of neutral and ionized hydrogen in galaxies 
can be estimated using 
the [OI]/Ha line ratio. The line ratio traces the neutral hydrogen in a general case of photoionization
irrelevant to specific photoionization source. In turn, one can estimate the ionized hydrogen mass 
or column density via the Balmer line emission \citep{osterbrock06}. 
We propose to combine this information obtained from the optical emission
line ratios to estimate the content of neutral hydrogen in galaxies.  

While radio observations provide a trusty and well-studied way for measuring HI, 
studying the spatial distribution of HI in galaxies can be done with just a few instruments 
in the world. In contrast, spatially resolved, optical spectroscopy can be performed 
with moderate size telescopes.
The progress in panoramic spectroscopy made last decades \citep{califa,sami,bundy15} allows us to study 
the spatial distribution, or maps of strong emission lines in large samples of galaxies. 

In this paper we use emission line maps obtained by the Mapping of Nearby Galaxies at APO survey
\citep[MaNGA,][]{bundy15} released in the frames of DR17 of the Sloan Digital Sky 
Survey \citep[SDSS,][]{abdurrouf22}.  The spectra maps allow us to estimate the gas-phase
metallicity, to assess the extinction in the galactic interstellar medium, and to make 
reasonable assumptions about the electron temperature. 
We develop the new method that allows one to estimate
the total mass of neutral hydrogen in galaxies as well as the radial distribution
of HI surface density. We obtain the HI mass and compare it with that estimated by 
traditional methods from radio observations. 

\section{Data and Sample Selection}
\subsection{The MaNGA sample}
\noindent
The MaNGA survey \citep{bundy15} employs the SDSS telescope \citep{gunn06} 
and R$\sim$2000 spectrographs \citep{smee13} to study 
spectra in the whole optical range 3,650 – 10,500\AA\, \citep{yan16} in more than 10,000 
nearby \citep[median redshift is 0.03,][]{wake17} galaxies. 
The Integral Field Unit (IFU) bundles provide spectra maps in up to 32 arcsec 
hexagonal areas \citep{drory15}. A significant fraction of these galaxies has been 
observed in the ALFALFA survey \citep{hoffman19,durbala20},  
which measured HI mass in the galaxies with a 3.5 arcmin angular resolution. 

\subsection{Structural Parameters and Inclination of MaNGA Galaixes}
\noindent
In order to compare the relatively high resolution optical and low resolution radio observations,  
we clean the sample of objects with reported potential confusion \citep{stark21} and of highly inclined galaxies.
We estimate the galaxy inclination via a Sersic fit isophote ellipticity from the 
NASA-Sloan Atlas, 
\citet{blanton11} (NSA\footnote{http://nsatlas.org}) catalog.
The r-band effective radius Re that we use for all galaxies is reported for MaNGA galaxies by 
the NSA catalog as well. 
We exclude highly-inclined galaxies (i $>$ 75\de) from the analysis of correlations with 
the ALFALFA sample (Figure \ref{fig1}) to avoid sampling highly reddened areas in galactic midplanes observed edge-on. 
Although the "confusion flag" reported for the radio observations \citep{stark21} should help reject potential 
nearby objects that may contribute to HI flux, 
we apply additional restrictions and remove objects with noticeable neighbours within the 
ALFALFA 3.5 arcmin aperture via a visual inspection of SDSS images. 
The resulting MaNGA-ALFALFA sample comprises 627 objects.

\section{Method and Results}
\noindent
As it was noticed by \citet{haffner09}, the first ionization potentials of O and H are similar to each other, 
which makes O$^+$/O ratio to be very close to H$^+$/H. This \citep{reynolds98} allows one to figure out 
the H$^+$/H$^0$ density ratio as 
$$N(H^0)/N(H^+) \,=\, 3.803 \times 10^{-5}\, I([OI])/I(H\alpha)\, C^{-1} / (O/H) ~~~ .$$ 
Here $(O/H)$ is the gas-phase metallicity,
I([OI])/I(\Ha) is the [OI]6300\AA\, emission line flux normalized by \Ha, and 
$$C \,=\, \zeta \, \frac{T_4^{1.85}}{1 + 0.605\, T_4^{1.105}}\, e^{-2.284/T_4} ~~~ ,$$ 
where $T_4$ = T/10,000K. Here the $\zeta$ is a constant of the order of unity \citep{reynolds98}. 
We assume that $\zeta$ =1 and $T_4$ = 0.7. Note that the estimated values depend on these constants 
nearly linearly, and we consider calibration of equation (1) via observational data below. 
From these equations above we estimate the neutral hydrogen mass in the MaNGA IFU bundles as
\begin{equation}
   log M(HI)_{est} = log M(HII) - 4.420 + log \frac{[OI]6300}{H\alpha} - log C - log \frac{O}{H}  
\end{equation}
In turn, \citet{osterbrock06} estimate  HII mass as
\begin{equation}
   M(HII) = m_p\, \sqrt{L(H\beta)\, S / e(H\beta)}  ~~, 
\end{equation}
where $m_p$ is the mass of proton, $e(H\beta) = 1.24 \times 10^{-25}\, T_4^{-0.88}$ ~ erg\, cm$^{-3}$\, s$^{-1}$ 
is the \Hb\, emissivity, L(\Hb) is the luminosity surface density in the \Hb\, line, and $S$ is the area 
covered by the IFU.  Additional refinements of this equation and the further discussion
can be found in \citet{revalski22}.

\citet{reynolds98} state that the method assumes that [OI] emission originates from 
collisional excitation by thermal electrons. The [OI]/\Ha~ ratio traces the 
HI content in warm ionized gas in regions where the collisional excitation
prevails. We expect that equation (1) will be correct in LI(N)ER and AGN regions
in galaxies, whereas in completely neutral HI as well as in the gas ionized by
shock waves it may fail.

\subsection{The Aperture Correction}

Since the apertures of spectroscopic MaNGA and radio ALFALFA observations 
are different, we have to apply an aperture correction. It is problematic 
in general, because the gas density distribution may not
follow the stellar surface density or brightness.
Fortunately, equations (1-2) allow us not only to obtain integral masses, but also
to estimate the HI mass in each spaxel and hence to analyze surface densities. 
In addition, useful relationships between global parameters of galaxies help
us figure out a reasonable aperture correction. Thus, \citet{broelis97,wang16} established clear
relationships between the total HI mass in galaxies with their "HI-size". The size
is defined as the galactic diameter at the HI surface density 
level $\Sigma_{HI}$ = 1 \Msunpc. As \citet{wang16,wang20} infer, the HI mass -- size relation
is very tight -- its r.m.s. scatter is 0.09 dex. Since our calibration sample of
ALFALFA galaxies has HI mass measurements, we can estimate the anticipated HI size. Note that
the HI gas is also detected at peripheries of galaxies beyond the "HI radius", 
but its surface density declines exponentially there \citep{wang16} and does 
not contribute significantly to the total HI mass in galaxies. 
 
Since our comparison sample has known HI masses, we start with the aperture correction 
based on the HI radius inferred from the total HI mass using corresponding, tight HI mass-size
relationship from \citet{wang16}. The possibility to derive HI mass based on optical spectroscopic data only,
when HI data are not available, motivates us to explore other methods of the aperture correction that do 
not use data of radio observations. 
Another approach to the HI-size estimation is an analysis of the surface density profiles
obtained via analogs of equations (1-2), which allows us to use the radial density distributions for 
inferring the aperture corrections.  We consider possible aperture correction methods
in \S3.4.


\subsection{Estimating the Integral HI Mass}

We combine equations (1) and (2) for the HI and HII mass with an aperture correction, 
and assume that $L(H\beta) = L(H\alpha) / 2.86$ \citep{osterbrock06} in the case of 
no extinction. Hereafter we use only 
$H\alpha$ emission for substituting to equation (2). 
Although the emission lines that we use have close wavelengths, we apply 
an extinction correction to all emission line fluxes 
based on observed $L(H\alpha) / L(H\beta)$ ratio
available from the MaNGA data and Cardelli et al extinction law \citep{cardelli89}.
Our aperture correction procedure is based 
on the HI mass $M_{HI}$ - HI size $D_{HI}$ relation 
$log\, D_{HI} \,=\, 0.506 \, log\, M_{HI} - 3.293$ by \citet{wang16}. The aperture
correction factor $f_{cor} = 1 + M_{out}/M_{in}$ is the ratio of the total HI mass
($M_{out}$ + $M_{in}$) within the $R_{HI} = 0.5\, D_{HI}$ limits to the HI mass within the 
radius limited by the IFU coverage $M_{in}$. 
We multiply the estimated HI masses by $f_{cor}$ to obtain the full masses. 
Note that the centres of most of MaNGA galaxies 
are allocated at the IFU centres.  

To estimate the aperture correction, we use the average 
HI surface density profiles collected by \citet{wang16}. 
Note that in this case the absolute 
value of HI surface density is not important for our method since we 
only compare the density distributions in the
inner and outer parts of galaxies. We assume that the HI density profile is approximated
by a third-order polynomial between the centre and $R_{HI}$.
We estimate $M_{out}$ and $M_{in}$ via analytic integration 
as $M_{in,out} = \int^{r_2}_{0} \Sigma_{HI}(r) 2\pi r dr $, where the 
radial integration limit
$r_{2}$ is chosen from the definition of $M_{in,out}$ given above.
In objects where 
the IFU coverage exceeds $R_{HI}$ (i.e. when we expect that all HI disk of a galaxy
completely falls in its IFU), we assume that $f_{cor}=1$. 

The gas-phase metallicity log(O/H) can be estimated from the emission line ratios in the galaxies as well. 
We consider three different methods of the abundance estimate in \S3.3 
below and find that the abundance
calibration based on a combination of [OIII] and [NII] lines 
(O3N2 hereafter, see its description in \S3.3) 
provides the best correspondence between the estimated and directly measured HI masses. 

\begin{figure}
\includegraphics[width=\columnwidth]{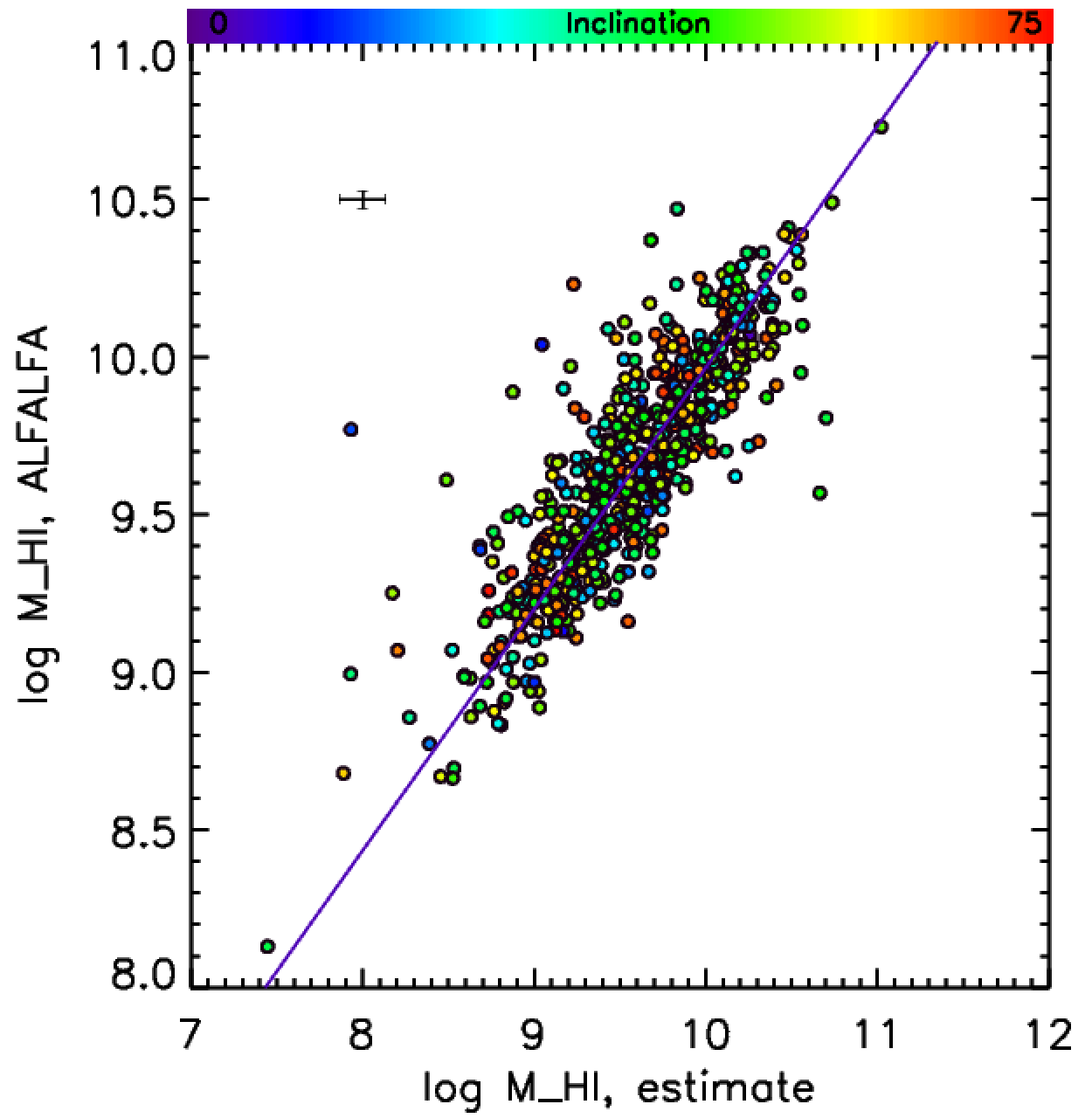}
\caption{Relation between the mass of neutral hydrogen estimated via equations (1-2) 
with the O3N2 metallicity calibration and HI mass from ALFALFA
\citep{hoffman19,durbala20} radio observations. The mean error bar that corresponds to 1-sigma uncertainties 
is shown in the upper left part of the plot. The blue solid line demonstrates the linear regression to the data
(see text).}
\label{fig1}
\end{figure}

Note that the correlation between the neutral and ionized hydrogen mass in the 
galaxies is significant and the Pearson correlation coefficient 
(hereafter CC) equals to 0.62. 
The estimated HI mass based on equations (1-2) essentially improves the 
correlation with the measured HI mass with respect to the HII: 
the Pearson correlation coefficient increases up to 0.91, see Figure \ref{fig1}. 
The empirical relation between the measured by ALFALFA HI mass and that estimated via 
equations (1-2) with the O3N2 metallicity calibration is  
\begin{equation}
 log M(HI)_{alfalfa} = (2.308 \pm 0.125) + (0.766 \pm 0.013)\, log M(HI)_{est} ,                      
\end{equation}
The rms scatter of the relationship in Figure \ref{fig1} is 0.149 dex.

\begin{figure}
\includegraphics[width=\columnwidth]{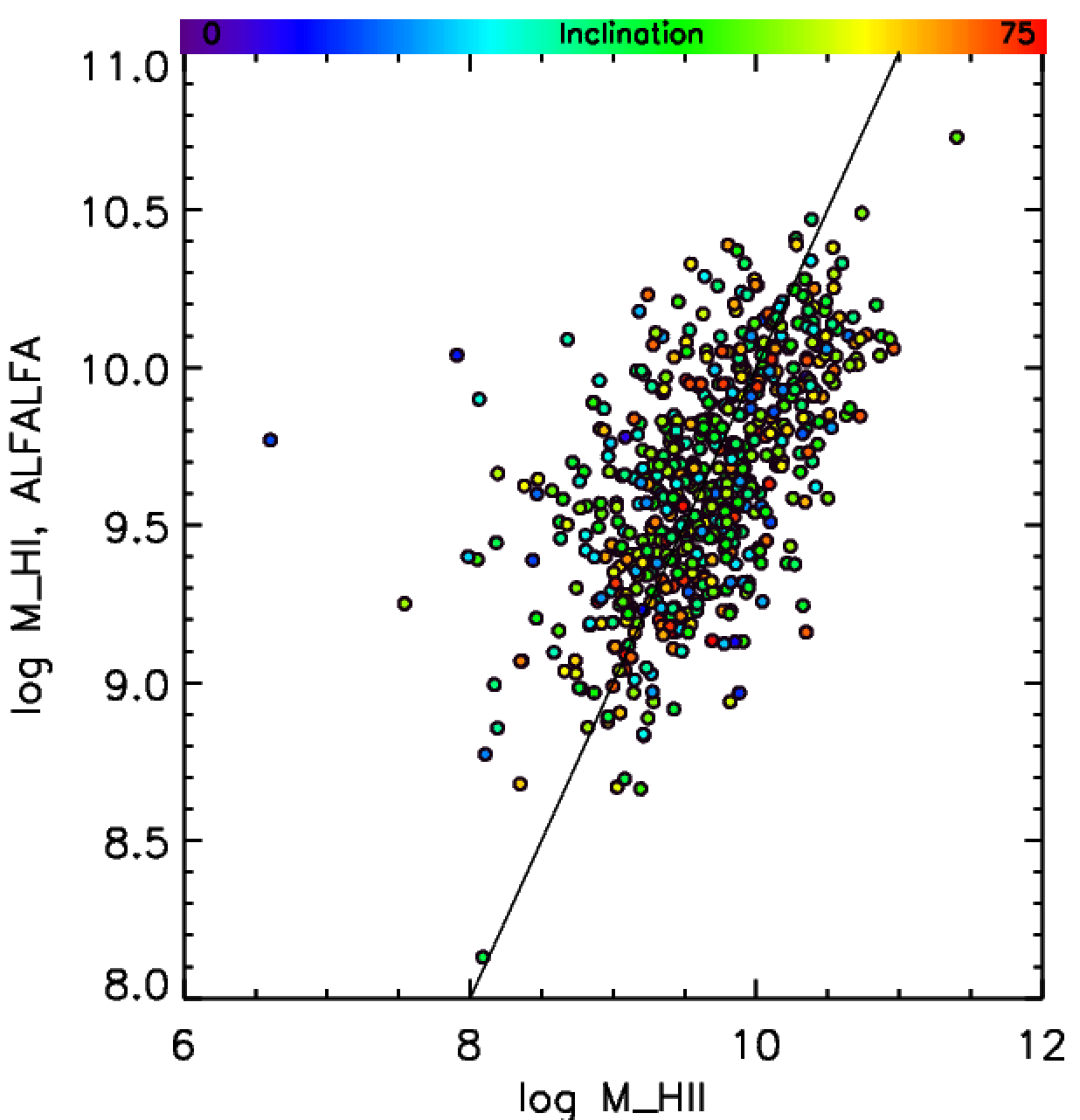}
\caption{Relation between the mass of ionized hydrogen estimated via equation (2) and
HI mass from ALFALFA \citep{hoffman19,durbala20} radio observations. The solid line 
indicates the one-by-one relation and is shown as a reference.}
\label{fig1b}
\end{figure}

We compare the HII and HI mass in Figure \ref{fig1b}. The HI mass is taken from same ALFALFA data set 
used for Figure \ref{fig1}.  

Equations (1-2) are motivated by physics, and we consider the tight correlation in Figure \ref{fig1} as an evidence 
of the reliable assessment of total HI mass in galaxies, given the limitations of the approach discussed later. 

\subsection{Metallicity Calibrations}
\noindent
We consider three calibrators to estimate the gas-phase metallicity for equation (1). 
The calibration proposed by \citet{dopita16} is based 
on red emission lines only: \Ha, [SII], and [NII] (HaNS). \citet{marino13} provide 
metallicity calibrations 
based on \Ha, [OIII], and [NII] (O3N2), or on \Ha\, and [NII] only (N2). 
The relations between the HI mass measured by ALFALFA and that estimated from the emission 
lines look similar, with the best one from the 
O3N2:   y= 2.308 + 0.766 x (it has been considered above), where
the y = log M(HI)$_{alfalfa}$ and x = log M(HI)$_{est}$. 
Its correlation coefficient CC = 0.913 and the rms = 0.149 dex. 
The N2 calibration yields y = 2.359 + 0.762 x, CC=0.910, rms = 0.154 dex, and the HaNS 
gives y = 2.582 + 0.756 x with CC = 0.873 and rms = 0.185 dex. 
Here the rms refers to the one-sigma scatter of points after 
one iteration of sigma-clipping with a 2-sigma limit. 

\subsection{Alternative Aperture Correction Methods}
\noindent
Although the aperture correction based on the HI mass-size relation works
well for the ALFALFA sample considered above, it requires preliminary 
knowledge or guess about the final HI mass, which makes sense only
for calibrating the relationship between the true and estimated 
HI mass. We need an independent way to estimate the aperture correction 
if our method is applied to an arbitrary galaxy with optical
spectroscopy but without radio observations. We check the relation 
between the galaxy optical scale, e.g. the effective radius 
$R_{eff}$ and the HI radius $R_{HI}$ and find that $R_{HI} = 3.82 \,+\, 3.69\, R_{eff}$
for the ALFALFA sample considered above. The relationship has CC = 0.75.
Our relationship is in agreement with HI-to-optical diameter ratio
estimated by \citet{bosma17}.

A more direct way to assess the HI size is to analyze the HI surface density
profiles and to estimate the $R_{HI}$ at the level of $\Sigma_{HI}$ = 1 \Msunpc.
We subdivide the galaxies by 7 elliptical annuli drawn according
to photometric ellipticity (b/a) of optical isophotes provided by the NSA
catalog. In this case the width of each annulus approximately matches the
spatial resolution of MaNGA data (2.5 arcsec FWHM). The HI surface density is obtained
via equations (1-2) as the mass in the elliptical annuli divided
by their area. 
We derive the HI mass in each spaxel instead of co-adding and re-analyzing
raw spectra in each annulus because the high-level MaNGA data are available  
in each spaxel, while re-running the MaNGA data reduction pipeline is more
difficult for regular MaNGA product users as well as for other relevant surveys 
explorers.

\begin{figure}
\includegraphics[width=\columnwidth]{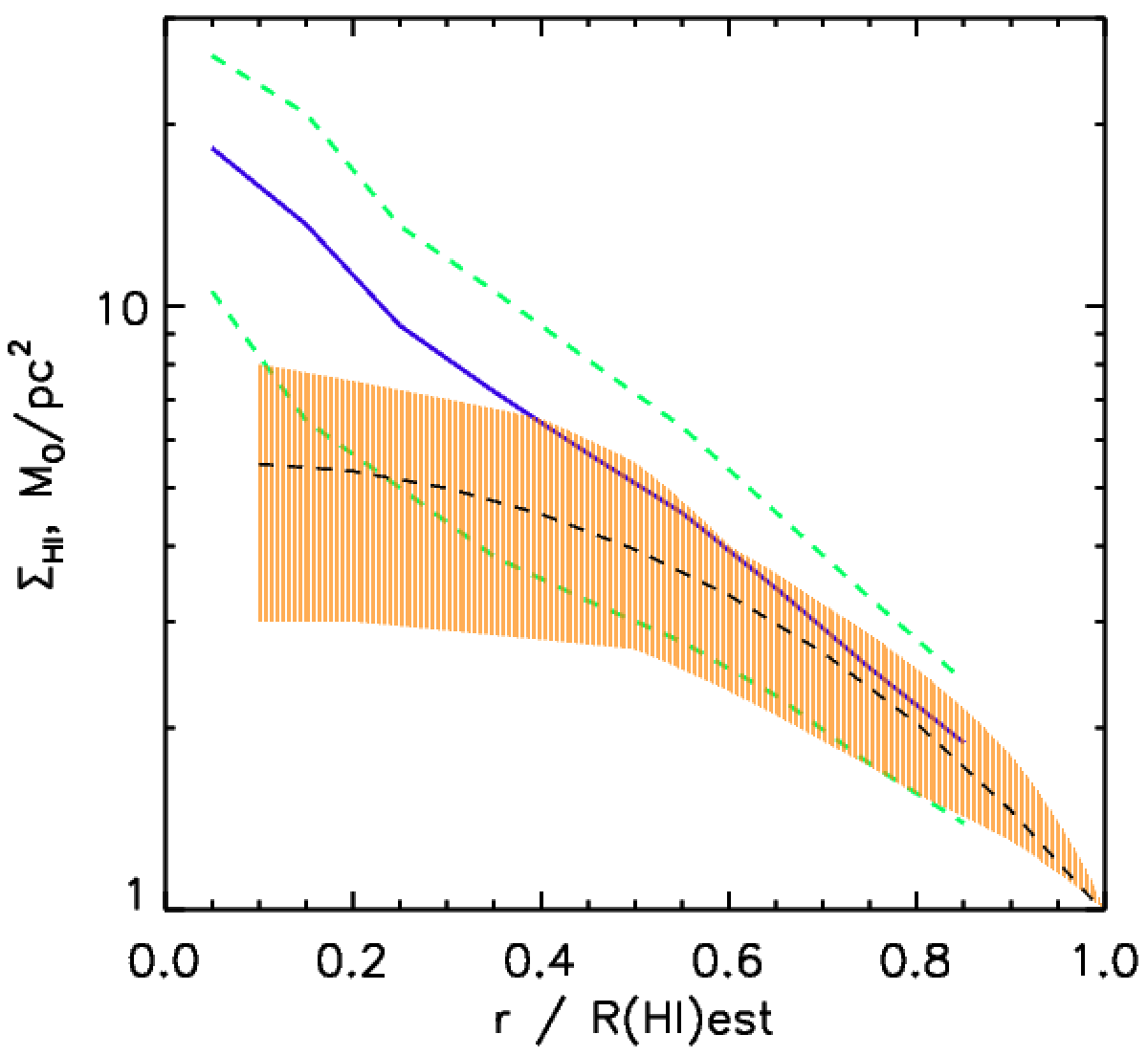}
\caption{Median radial distribution of HI surface density (solid blue curve) made
with our ALFALFA sample. The Median Absolute Deviation curves are shown
with dashed green lines. The radius is normalized by $R_{HI}$ obtained by the extrapolation
of estimated gas surface density curves, see text.
The average surface density profiles considered by \citet{wang16} for the 
galaxies with resolved radio observations are shown by the shaded brown area.
The dashed black curve demonstrates our polynomial fitting to 
the radial profiles from \citet{wang16}.
}
\label{fig2}
\end{figure}

The blue solid curve in Figure \ref{fig2} shows the median radial distribution of HI surface
density for the ALFALFA sample. The green dashed curves designate 
one Median Absolute Deviation from the median curve.
The shaded brown area shows the average HI surface density profiles
from \citet{wang16}. Our polynomial fitting is designated by the
dashed black curve. 
Our median profile overlaps the
surface density profiles of the galaxies with resolved radio observations.
They demonstrate significant difference in central regions, but have good agreement at the
periphery.

Most of $\Sigma_{HI}$ distributions estimated by us for ALFALFA sample
do not extend to the level of 1 \Msunpc. At the same time, most of the 
surface density profiles look linear in the $log\, \Sigma_{HI}$ --
radius plots, which allows us to find $R_{HI}$ via the linear 
extrapolation using the last 3 data points along the radius.
The obtained $R_{HI}$ and the $\Sigma_{HI}$ radial distributions are 
then used to calculate $M_{in}$ and $M_{out}$ (see above), and therefore to 
estimate the aperture correction $f_{cor}$.
When this method is applied with equations (1-2) and  
O3N2 metallicity calibration, it results to the relation
\begin{equation}
  log M(HI)_{alfalfa} = (2.222 \pm 0.184) + (0.810 \pm 0.020)\, log M(HI)_{est} ,                      
\end{equation}
with CC = 0.802 and rms = 0.229.

We consider one more approach of the aperture correction and employ the universal slope
of the HI surface density profile at large radii reported by \citet{wang16}. We assume
that all our surface density profiles get this exponential slope of 0.2$R_{HI}$
beyond the last estimated surface density data points on our radial profiles.
In this case the correlation between $log M(HI)_{alfalfa}$ and $log M(HI)_{est}$ 
slightly improves to CC = 0.814.

\subsection{Testing Secondary Calibration of the HII-HI Mass Difference}
\noindent
While the connection between HI and HII mass 
considered in \S3 is physically motivated, it scales the HII mass
with a multiplicative coefficient, which can be
generalized as an empirical coefficient in a more wide approach.
In this chapter we decided to make reasonable assumptions of additional
factors that could serve as indicators of the gas state in the interstellar medium.

Here we check if we could utilize some indicators of the radiation field status
as secondary calibrators of the difference between the estimated ionized gas mass 
and directly measured HI mass
in order to improve the HII-HI mass correlation by parameterizing the mass 
difference with the mean, flux weighted ratio of those spectra line ratios in the galaxies. 
As possible indicators we consider
the excitation parameter [OIII]/[OII] \citep{pilyugin16}, 
ionization sensitive ratio [SIII]/[SII] \citep{kewley02}, 
and the radiation softness $\eta$ \citep{vilchez88}.
We find that these indicators 
correlate with the HII-HI mass ratio with CC$\sim$0.5 or less. 
Applying the secondary parameterization to the (HII - HI) mass ratio slightly improves the HII-HI mass relation, 
but the resulting estimated HI mass does not correlate with the measured HI mass as well as for
the more direct method based on equations (1-3) considered above.

\subsection{Testing Predictions with the HIMANGA Sample}
\noindent
In order to test the HI mass estimation via equations (1-2), we use an independent sample of MaNGA galaxies 
for which single-dish radio observations were performed in the frames of HIMANGA project \citep{masters19,stark21}. 
The HIMANGA sample was observed with the Green Bank radio telescope \citep[GBT, ][]{masters19}, and it 
provides the HI mass 
with a 9 arcmin angular resolution for a large fraction of MaNGA galaxies \citep{masters19,stark21}. 
HIMANGA program observed more than a half of 
all MaNGA galaxies, but not in all of the objects HI was detected.  
     
\begin{figure}
\includegraphics[width=\columnwidth]{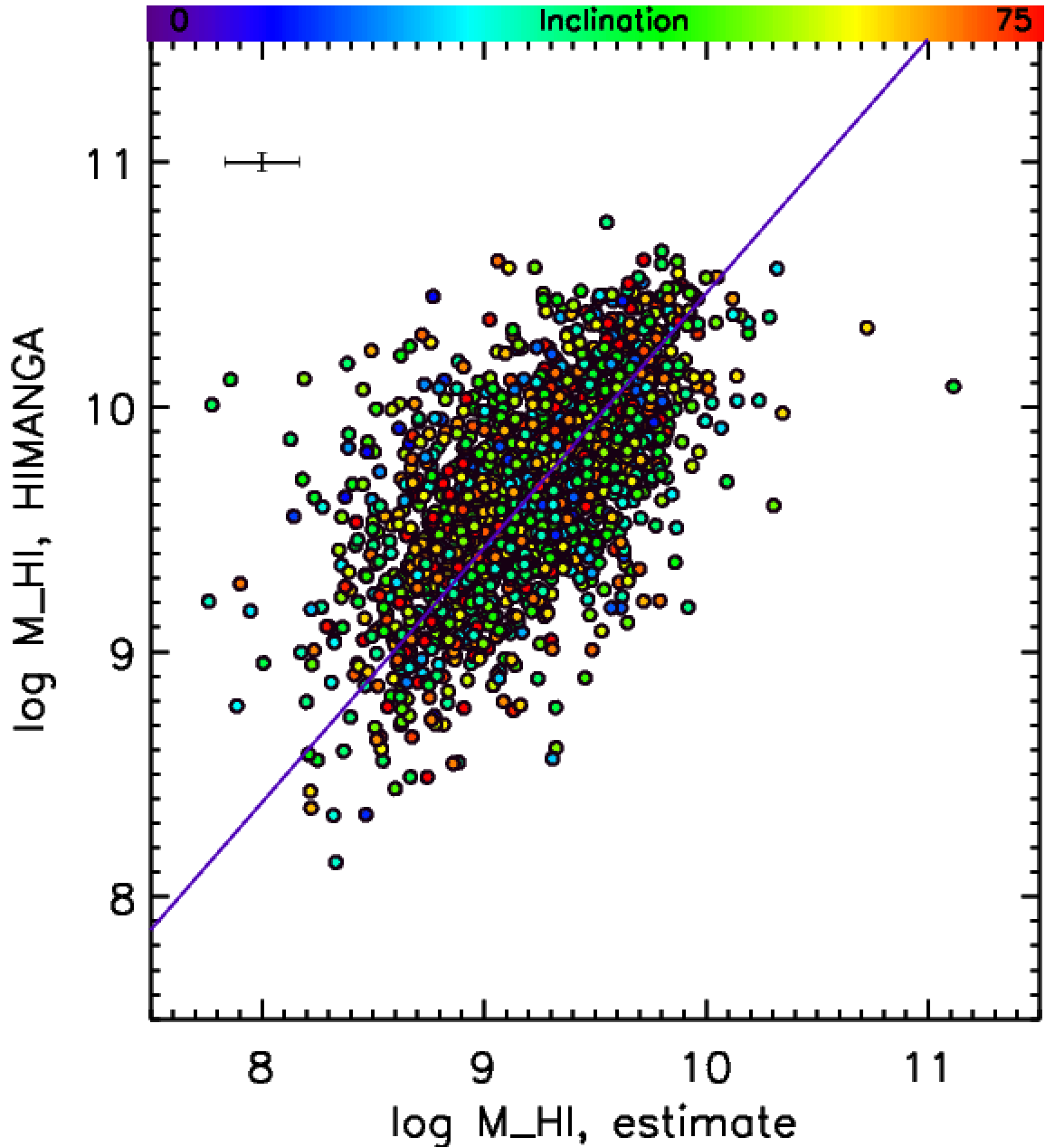}
\caption{Relation between the mass of neutral hydrogen estimated via equations (1-2) and calibrated with 
equation (4) and the HI mass from HIMANGA/GBT radio observations \citep{masters19,stark21}. 
The mean error bar that corresponds to 1-sigma uncertainties is shown in the upper left part of the plot.}
\label{fig3}
\end{figure}

We apply our equations (1-2, 4) to all HIMANGA galaxies with detected HI and with good data 
reduction flags \citep{masters19}. The sample comprises 1811 objects. 
The aperture correction is obtained from the estimated HI
surface density radial distributions. 
This independent sample demonstrates a tight, one-to-one correlation between the estimated 
(via equations 1-2, 4) and measured by GBT mass of neutral hydrogen: 
\begin{equation}
log M(HI_{GBT}) = (0.070 \pm 0.161) + (1.039 \pm 0.017)\, x \, log M(HI)_{est} ,  
\end{equation}
see Figure \ref{fig3}. The CC = 0.736 and rms = 0.291. We also tried to apply the calibration
equation (3) instead of equation (4), and obtained similar relationship 
with slightly different coefficients, and similar rms (0.291) and CC (0.740). 
If we estimate the correction factor using the $R_{HI} - R_{eff}$ relation
instead, the correlation and rms stay at the same level.

The correlation for the HIMANGA sample is worse with respect to ALFALFA. We 
attribute it to the worse angular resolution of HIMANGA observations. Since the area of 
the HIMANGA beam is by the order of magnitude larger, more companion galaxies can contribute
to the HI flux and hence to overestimate the HI mass in the HIMANGA aperture in comparison with
the optical spectroscopy.  The aperture correction calculations are less reliable
for the much larger HIMANGA beam area.

The method of HI mass measurement that we propose here is based solely on optical spectra observations. 
If the N2 abundance calibration \citep{marino13} is used, it would require observations in 3 emission lines 
only ([OI], [NII], and \Ha) from the red part of optical spectra. These data can be obtained
via e.g. narrow-band imaging for multiple galaxies at the same time. 

This method can be applied to study the distribution of surface density 
in diffuse gas in galaxies \citep{reynolds98}, not only to assess the 
total hydrogen mass. 
In this case the spatial resolution of observations will be much 
better than that for single dish radio observations.  

\subsection{Applying the Mass -- Metallicity Relations}
\noindent
While MaNGA provides a way to estimate the gas-phase metallicity in each 
spaxel or resolution element in the galaxies, we envision a survey that would
benefit from using as few emission lines as possible. In this case we attempt
to substitute the metallicity measurements across the galaxies made with
different metallicity calibrators by an integral galactic metallicity estimated
from correlations between global galactic parameters. We substitute
the stellar mass -- gas phase metallicity relation figured out by 
\citet{tremonti04} to equation (1), estimate the HI mass with equations 
(1-2, 4) and compare it with the true HI masses for the HIMANGA sample,
similar to what has been done for equation (4). 
In this case the relationship looks like 
$$ log M(HI_{GBT}) = (0.134 \pm 0.164) + (1.066 \pm 0.018)\, log M(HI)_{est} , $$ 
with CC = 0.718 and rms = 0.306.

\section{Limitations of the Approach}
\noindent
The proposed method has its limitations. 
It should work in the cases when the intergalactic medium contains ionized and 
neutral hydrogen in a mix, where the collisional excitation prevails. 
In turn, it may fail for the purely ionized gas, like inside the Stromgren spheres \citep{osterbrock06}
around hot stars. 
This effect may introduce systematics between high- and low-mass galaxies,
and it can be explored in detail with fine resolution (dozens of parsecs)
radio and IFU maps in future studies. 

It is difficult to find a large sample 
of galaxies with superior, arcsecond spatial resolution radio observations and with 
full coverage by panoramic spectroscopy at the same time, 
so we have to correct for the aperture difference. In \S3.4 we apply empirical
relations in order to estimate the HI size in galaxies via their optical parameters 
instead of applying the tight HI mass- HI size relation known from radio data. 
It increases the scatter of the estimated vs measured mass relation for 
the price of a possibility to figure out HI masses and surface densities 
from optical spectroscopy only. 
It indicates that the empirical HI size uncertainty together with the
physical limitations mentioned above can contribute to the slope deviation 
from unity in Figure \ref{fig1} and to overall scatter of 
the "estimated vs directly measured" relation.


\section{Conclusions}
\noindent
In this study we propose a method of estimating the neutral hydrogen HI mass 
from optical spectroscopic observations.
The method is motivated by the physical relations between the strength of ionized gas line-emission 
and the gas ionization fraction. 
We calibrate the method using HI masses measured by the ALFALFA survey. The scatter
of the (estimated - observed) mass difference is of the order of 0.15 dex, which is significantly
less than 0.3-0.4 dex that can be obtained with other, purely empirical calibrations
\citep{kannappan04,zhang09,li10,eckert15} of the total HI mass in galaxies. 
The proposed method is verified via an independent sample of HI measurements
provided by the HIMANGA project.

The successful prediction of total galactic HI mass and HI surface density radial distributions
via this method opens possibilities for massive studies of HI content in galaxies with detected 
strong emission lines, presumably in a wide range of redshifts. 

\section*{Acknowledgements}
\noindent
We'd like to thank David Stark for productive discussions about
the subject of the paper. We thank the anonymous referee whose comments and 
suggestions significantly improved the paper.  The study is partly supported
by RSCF grant 22-12-00080.

The study makes use of the SDSS-IV MaNGA data available from  http://www.sdss.org/dr17/.
Funding for the Sloan Digital Sky Survey IV has been provided by the 
Alfred P. Sloan Foundation, the U.S. Department of Energy Office of 
Science, and the Participating Institutions. 

SDSS-IV acknowledges support and resources from the Center for High Performance Computing  at the 
University of Utah. The SDSS-IV website is www.sdss4.org.

SDSS-IV is managed by the Astrophysical Research Consortium 
for the Participating Institutions of the SDSS Collaboration including 
the Brazilian Participation Group, the Carnegie Institution for Science, 
Carnegie Mellon University, Center for Astrophysics | Harvard \& 
Smithsonian, the Chilean Participation Group, the French Participation Group, 
Instituto de Astrof\'isica de Canarias, The Johns Hopkins 
University, Kavli Institute for the Physics and Mathematics of the 
Universe (IPMU) / University of Tokyo, the Korean Participation Group, 
Lawrence Berkeley National Laboratory, Leibniz Institut f\"ur Astrophysik 
Potsdam (AIP),  Max-Planck-Institut f\"ur Astronomie (MPIA Heidelberg), 
Max-Planck-Institut f\"ur Astrophysik (MPA Garching), 
Max-Planck-Institut f\"ur Extraterrestrische Physik (MPE), 
National Astronomical Observatories of China, New Mexico State University, 
New York University, University of Notre Dame, Observat\'ario 
Nacional / MCTI, The Ohio State University, Pennsylvania State 
University, Shanghai Astronomical Observatory, United 
Kingdom Participation Group, Universidad Nacional Aut\'onoma 
de M\'exico, University of Arizona, University of Colorado Boulder, 
University of Oxford, University of Portsmouth, University of Utah, 
University of Virginia, University of Washington, University of 
Wisconsin, Vanderbilt University, and Yale University.

\section*{Data Availability}
This work makes use of SDSS/MaNGA project data publicly available at
https://www.sdss4.org/dr17/data\_access/.

{}

\bsp	
\label{lastpage}

\end{document}